# Tracking the affordability of least-cost healthy diets helps guide intervention for food security and improved nutrition

**Policy Comment submission for *Food Policy*'s 50th anniversary special issue, reflecting on the impact of previous papers in the journal**

Submitted 29 March 2025, this version revised 23 June 2025

**William A. Masters**
Friedman School of Nutrition Science and Policy
and Department of Economics, Tufts University
william.masters@tufts.edu

**Highlights**

- Bai et al. (2020) calculated least-cost nutrient adequate diets at global scale
- Bai et al. (2021) reviewed global retail food price data available for diet costing
- Other papers published in *Food Policy* accelerated use of least-cost diets globally
- Comparing benchmark diets to observed food choices can guide policy
- Results point to lower-cost options, safety nets and displacement by other foods

**Abstract**

This Policy Comment describes how the *Food Policy* article entitled "Cost and affordability of nutritious diets at retail prices: Evidence from 177 countries" (first published October 2020) and "Retail consumer price data reveal gaps and opportunities to monitor food systems for nutrition" (first published September 2021) advanced the use of least-cost benchmark diets to monitor and improve food security. Those papers contributed to the worldwide use of least-cost diets as a new diagnostic indicator of food access, helping to distinguish among causes of poor diet quality related to high prices, low incomes, or displacement by other food options, thereby guiding intervention toward universal access to healthy diets.

*This work was conducted as part of the Food Prices for Nutrition project (INV-016158), funded by the Bill & Melinda Gates Foundation and the Foreign, Commonwealth and Development Office of the United Kingdom. Project news is here: https://www.linkedin.com/showcase/food-prices-for-nutrition, and details of the project are here: https://sites.tufts.edu/foodpricesfornutrition.*

**Tracking the affordability of least-cost healthy diets helps guide intervention for food security and improved nutrition**

In October of 2020, *Food Policy* published "Cost and affordability of nutritious diets at retail prices: Evidence from 177 countries" (Bai et al., 2020).  That paper demonstrated how and why least-cost nutrient adequate diets could be used to measure food access, using standardized item prices matched to food composition and requirements for dietary energy plus lower and upper bounds for 21 essential nutrients. The study's results were also published in the UN agencies' SOFI 2020 report, *State of Food Security and Nutrition in the World: Transforming Food Systems for Affordable Healthy Diets* (FAO, IFAD, UNICEF, WFP and WHO, 2020) alongside the simplified Cost and Affordability of Healthy Diets (CoAHD) indicator using dietary energy plus food group targets specified in national dietary guidelines (Herforth et al. 2020). This work pioneered the use of benchmark least-cost diets as a diagnostic indicator of access to sufficient quantities of foods needed for an active and healthy life around the world.

Then in September 2021, a *Food Policy* review article entitled "Retail consumer price data reveal gaps and opportunities to monitor food systems for nutrition" (Bai et al., 2021) showed how existing data collected for inflation monitoring, market information and early warning systems could be adapted for measuring access to healthy diets, in ways that were then introduced in a variety of countries worldwide (Herforth et al. 2024). The first country to publish least-cost healthy diets as an official national statistic was Nigeria (National Bureau of Statistics 2024), where the metric soon proved useful in policy debates as a benchmark for what national minimum wages should be able to buy (Muhammad, 2024). Nigerian analysts also use these data to advocate for change in agricultural and trade policy (Adio 2024), and to address regional differences in access to healthy diets (Ejiade and Olorunfemi (2025).

Before 2020, the use of optimization models to calculate which foods would meet human needs at low cost was done primarily to guide nutrition assistance, for example in the U.S. Thrifty Food Plan (Wilde and Llobrera 2009) or in low income countries through the World Food Program's "Fill the Nutrient Gap" initiatives (de Pee et al. 2017). The concept had been introduced by Stigler (1945) and was used primarily to compute specific meal plans for particular populations, with a rare use as a price index over time being the doctoral work of Patricia O'Brien-Place (1983) comparing inflation for least-cost diets to the consumer price index in the United States, and use across countries as the food component of an absolute poverty line (Allen 2017).

The new work on diet cost and affordability introduced in *Food Policy* and other outlets involved calculation of least-cost adequate diets at global scale, identifying the millions of food combinations that could meet human nutritional requirements using locally available items at each place and time as described by Masters et al. (2025a, 2025b). Several important papers in *Food Policy* added to this literature, such as Raghunathan et al. (2020) on cost and affordability of healthy diets in India, Schneider (2022) on cost and affordability of nutrient-adequate diets in Malawi, Headey et al. (2024) on methods used for global monitoring of cost), and affordability of healthy diets, and most recently Stehl et al. (2025) on how global poverty measurement relates to the cost and affordability of healthy diets. Rapid growth in use of these concepts around the world is shown by the relative frequency of related phrases in English-language books and policy documents reported in Figure 1 below.

Figure 1. Frequency of selected phrases related to diet costs in books and documents, 2010-2022

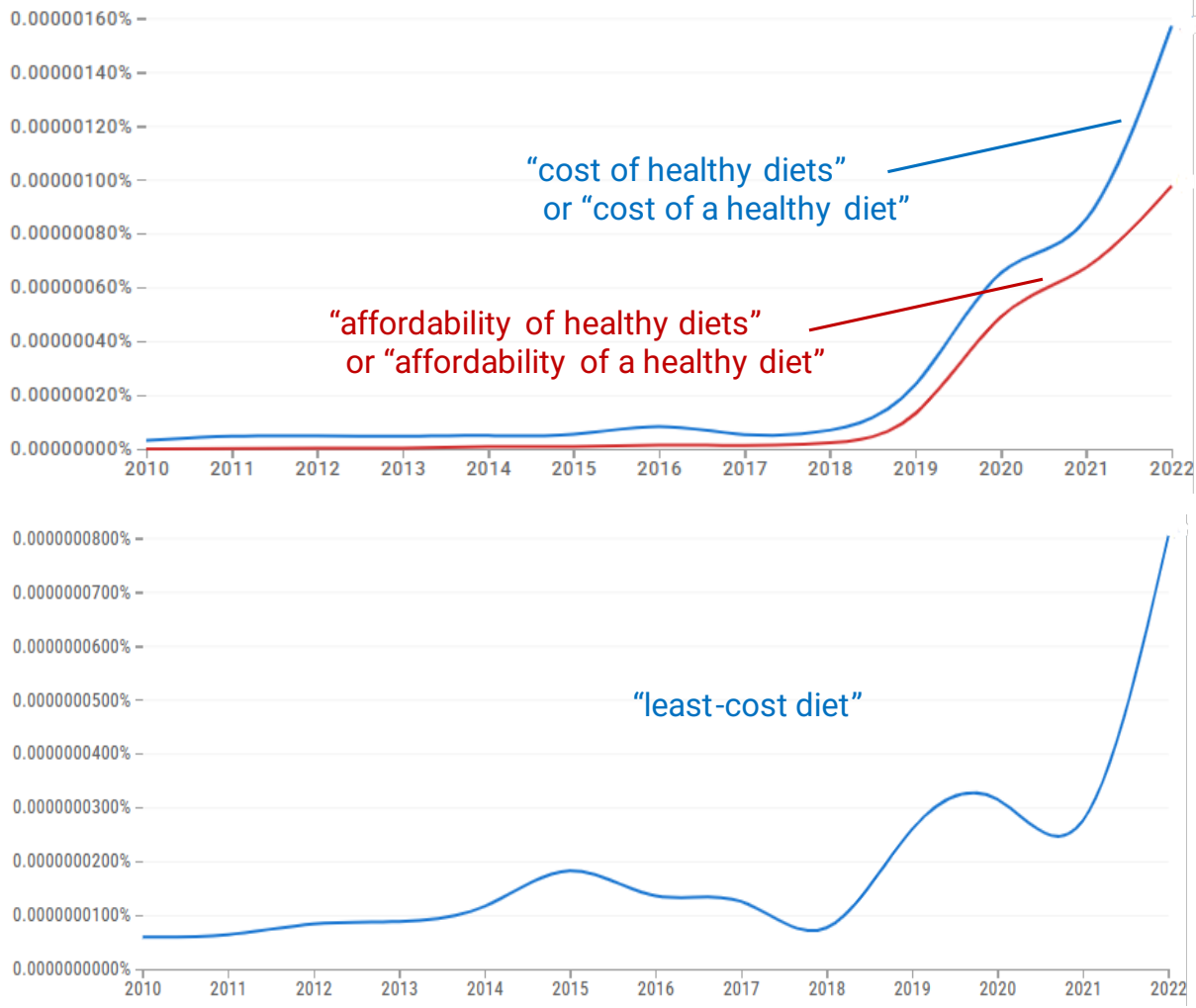

Source: Author's calculation from Google Ngram viewer. Data shown are the phrase indicated as a proportion of all phrases with that number of words, from https://books.google.com/ngrams:
https://books.google.com/ngrams/graph?content=least-cost+diet&year_start=2010&year_end=2022&corpus=en&smoothing=0, and
https://books.google.com/ngrams/graph?content=%28cost+of+healthy+diets+%2B+cost+of+a+healthy+diet%29%2C%28affordability
+of+healthy+diets+%2B+affordability+of+a+healthy+diet%29&year_start=2010&year_end=2022&corpus=en&smoothing=0

Tracking the affordability of least-cost healthy diets provides a new way of measuring food security in terms of access to nutritious options, designed as a diagnostic indicator to distinguish between three possible causes of malnutrition:

   **(a) High prices** or unavailability of even the lowest-priced foods to meet each nutritional requirement, calling for investment and innovation to lower costs of production and distribution along the supply chain to reduce retail prices;

   **(b) Low incomes** available for food, calling for improved livelihoods and safety nets to increase total income and limit required spending on nonfood needs, increasing ability to buy a balanced diet for lifelong health;

(c) **Displacement** of low-cost healthy options by more expensive and often less nutritious options, for reasons such as taste and preferences, time use and cooking cost, or aspirations shaped by marketing efforts.

For policy analysis, benchmark least-cost healthy diets have been used in annual global food security monitoring through the UN system's flagship SOFI reports and the FAOSTAT database ever since the initial publication of CoAHD in 2020. Since 2022, the CoAHD has been published jointly by FAO and the World Bank, using least cost items meeting global targets for energy balance and diversification across Healthy Diet Basket targets for six food groups derived from commonalities among national dietary guidelines (Herforth et al. 2022). The CoAHD method is helpful because it identifies least cost options using rank order selection instead of linear programming, and because meeting food group targets leads to dietary patterns that are aligned with national policies while also coming very close to nutrient adequacy (Herforth et al. 2025). The resulting diet costs for each country are published each year simultaneously by FAO (2025) and the World Bank (2025), then used in policy advice for example by comparing least-cost healthy diets to projected consumption patterns under various scenarios in simulation models (Sanchez and Cicowiez 2023, Sanchez et al. 2024), or comparing prices with and without trade restrictions (Gilbert et al. 2024) and targeting nutrition assistance (Wallingford et al. 2024).

Each use of the new methods compares benchmark least-cost diets to other data. Comparing diet costs over space and time reveals where, when and how each food system does or doesn't deliver cost-effective options for a healthy diet . This could be especially important to identify seasonal or spatial bottlenecks in supply for particular types of food where prices are unusually high, and identify the investments and innovations needed to make at least some low-cost options available everywhere. Comparing diet costs to income then reveals where, when and how each population's total spending is insufficient to obtain a healthy diet. Unaffordability might be caused by low household income, calling for safety nets and improved employment options, but might also be caused by high cost of housing or other nonfood requirements. Finally, comparing benchmark diets to foods actually consumed reveals where, when and how the least expensive options for a healthy diet are displaced by higher-priced or less nutritious items for a variety of reasons. Even consumers who seek a low-cost healthy diet would not know which items to use in what proportions, and most consumers have other objectives such as taste, convenience, and aspirations shaped by marketing for various kinds of food.

In conclusion, Bai et al. (2020, 2021) helped demonstrate how retail price data already being collected around the world could be used to calculate benchmark least-cost diets for measuring food access. For policy purposes, upper and lower bounds for nutrient adequacy are often replaced by energy targets for each food group as in the EAT-Lancet diet used for Hirvonen et al. (2019), or the national dietary guidelines and Healthy Diet Basket targets used in the CoAHD metric developed by Herforth et al. (2020, 2022, 2025). The word frequency data in Figure 1 and widespread adoption of these methods reveals the central role of *Food Policy* in publishing frontier work on global food security and nutrition, in this case towards universal access and use of healthy diets around the world.

# References

Adio, W. (2024), Urgent Need to Halt Soaring Food Prices. *This Day* (Lagos), Backpage Essay. https://www.thisdaylive.com/2024/02/04/urgent-need-to-halt-soaring-food-prices

Allen, R.C. (2017). Absolute Poverty: When Necessity Displaces Desire. *American Economic Review* 107 (12): 3690–3721. https://doi.org/10.1257/aer.20161080.

Bai, Y., R. Alemu, S.A. Block, D. Headey, and W.A. Masters (2020). Cost and affordability of nutritious diets at retail prices: Evidence from 177 countries. *Food Policy* 99:101983. https://doi.org/10.1016/j.foodpol.2020.101983.

Bai, Y., L. Costlow, A. Ebel, S. Laves, Y. Ueda, N. Volin, M. Zamek, A. Herforth, and W.A. Masters (2021). Retail consumer price data reveal gaps and opportunities to monitor food systems for nutrition. *Food Policy* 104: 102148. https://doi.org/10.1016/j.foodpol.2021.102148

de Pee, S., G. Baldi, I. Bose, and L. Kiess (2017). 'Fill the Nutrient Gap' approach to situation analysis and decision-making for improving nutrition. Presented at the 21st International Congress of Nutrition, abstract in *Annals of Nutrition and Metabolism* 71(suppl. 2): 251. https://doi.org/10.1159/000480486

Ejiade, O. and O. Olorunfemi (2025). The rising cost of a healthy diet in Southwest Nigeria: A ticking time bomb for public health. Ibadan, Nigeria: Development Agenda for Western Nigeria (DAWN) Commission. https://dawncommission.org/the-rising-cost-of-a-healthy-diet-in-southwest-nigeria

FAO (2025), Cost and Affordability of a Healthy Diet database. Rome, FAO. https://www.fao.org/faostat/en/#data/CAHD.

FAO, IFAD, UNICEF, WFP and WHO (2020). *The State of Food Security and Nutrition in the World 2020: Transforming food systems for affordable healthy diets*. Rome: FAO. https://doi.org/10.4060/ca9692en.

O'Brien-Place, P. (1983). Impact of inflation on least-cost diets in the United States. Doctoral dissertation, Cornell University. https://catalog.library.cornell.edu/catalog/779424

Headey, D.D., K. Hirvonen, and H. Alderman (2024). "Estimating the Cost and Affordability of Healthy Diets: How Much Do Methods Matter?" *Food Policy* 126:102654. https://doi.org/10.1016/j.foodpol.2024.102654.

Herforth, A., Bai, Y., Venkat, A., Mahrt, K., Ebel, A., Masters, W.A. (2020). Cost and affordability of healthy diets across and within countries. Background paper for The State of Food Security and Nutrition in the World 2020. FAO Agricultural Development Economics Technical Study No 9. https://doi.org/10.4060/cb2431en.

Herforth, A.W., A. Venkat, Y. Bai, L. Costlow, C. Holleman, & W.A. Masters (2022). Methods and options to monitor the cost and affordability of a healthy diet globally. Background paper for The State of Food Security and Nutrition in the World 2022. FAO Agricultural Development Economics Working Paper 22-03. Rome, FAO. https://doi.org/10.4060/cc1169en.

Herforth, A.W., R. Gilbert, K. Sokourenko, T. Fatima, O. Adeyemi, D. Alemayehu, E. Arhin, F. Bachewe, Y. Bai, I. Chiosa, T. Genye, H. Haile, R. Jahangeer, J. Kinabo, F. Mishili, C.D. Nnabugwu, J. Nortey, B. Ofosu-Baadu, A. Onabolu, D. Sarpong, M. Tessema, D. TT. Van, A. Venkat and W.A. Masters (2024). Monitoring the Cost and Affordability of a Healthy Diet within countries: Building systems in Ethiopia, Ghana, Malawi, Nigeria, Pakistan, Tanzania, and Viet Nam. *Current Developments in Nutrition*, 8: e104441. https://doi.org/10.1016/j.cdnut.2024.104441.

Herforth, A.W., Bai, Y., Venkat, A., Masters, W.A. (2025). The Healthy Diet Basket is a valid global standard that highlights lack of access to healthy and sustainable diets. *Nature Food* 6: 622–631. https://doi.org/10.1038/s43016-025-01177-0.

Hirvonen, K., Y. Bai, D.D. Headey, and W.A. Masters (2020). "Affordability of the EAT–Lancet Reference Diet: A Global Analysis." *The Lancet Global Health* 8 (1): e59–66. https://doi.org/10.1016/S2214-109X(19)30447-4.

Gilbert, R., L. Costlow, J. Matteson, J. Rauschendorfer, E. Krivonos, S.A. Block, and W.A. Masters (2024). Trade policy reform, retail food prices and access to healthy diets worldwide. *World Development* 177 (2024): 106535. https://doi.org/10.1016/j.worlddev.2024.106535

Masters, W.A., J.K. Wallingford, R.D. Gilbert, E.M. Martinez, Y. Bai, K. Sokourenko, and A.W. Herforth (2025a). Are healthy foods affordable? The past, present, and future of measuring food access using least-cost diets. *Annual Review of Resource Economics* 17: forthcoming.

Masters, W.A., J.K. Wallingford, A.W. Herforth and Y. Bai (2025b). Measuring food access as affordability of least-cost healthy diets worldwide. *Agricultural Economics*, 56(3): 360-372. https://doi.org/10.1111/agec.70028

Muhammad, Abdullahi (2024). N70,000 minimum wage: Protecting the inconsequential majority. Abuja, Nigeria: Blueprint Newspaper. https://blueprint.ng/n70000-minimum-wage-protecting-the-inconsequential-majority

National Bureau of Statistics (2024). "Cost of Healthy Diets Monthly Reports." Abuja, Nigeria: NBS. nigerianstat.gov.ng/elibrary?queries=cost+of+healthy+diet, and https://microdata.nigerianstat.gov.ng/index.php/catalog/146

Raghunathan, K., D. Headey and A.W. Herforth (2020), Affordability of nutritious diets in rural India. *Food Policy* 99: 101982. https://doi.org/10.1016/j.foodpol.2020.101982

Sánchez, M.V. & M. Cicowiez (2023). Optimal allocation of agriculture's public budget can improve transformation and healthy diets access in Ethiopia. *Journal of Policy Modelling* 45(6): 1262-1280. https://doi.org/10.1016/j.jpolmod.2023.09.005

Sánchez, M.V., M. Cicowiez, V. Pernechele and L. Battaglia (2024). The opportunity cost of not repurposing public expenditure in food and agriculture in sub Saharan African countries. FAO Agricultural Development Economics Working Paper 24-07. Rome, FAO. https://doi.org/10.4060/cd3753en

Schneider, K.R. (2022). Nationally representative estimates of the cost of adequate diets, nutrient level drivers, and policy options for households in rural Malawi. *Food Policy* 113: 102275. https://doi.org/10.1016/j.foodpol.2022.102275

Stehl, J., L. Depenbusch, and S. Vollmer (2025). Global poverty and the cost of a healthy diet. *Food Policy* 132: 102849. https://doi.org/10.1016/j.foodpol.2025.102849

Wallingford, J., S. de Pee, A.W. Herforth, S. Kuri, Y. Bai and W.A. Masters (2024). Measuring food access using least-cost diets: Results for global monitoring and targeting of interventions to improve food security, nutrition and health. *Global Food Security*, 41: 100771. https://doi.org/10.1016/j.gfs.2024.100771

Wilde, P. and J. Llobrera (2009). Using the Thrifty Food Plan to Assess the Cost of a Nutritious Diet. *Journal of Consumer Affairs*, 43(2): 274-304. https://doi.org/10.1111/j.1745-6606.2009.01140.x

World Bank (2024), Food Prices for Nutrition database. Washington, DC: The World Bank. https://doi.org/10.57966/41AN-KY81